\documentclass[preprint,amsmath,amssymb,aps]{revtex4}

\usepackage{graphicx}
\usepackage{dcolumn}
\usepackage{bm}
\newcommand{\degree}{\ensuremath{^\circ}}

\begin{document}
\title{Monitoring the Bragg peak location of 73 MeV/u carbon ions by means of prompt $\gamma$-ray measurements}
\author{E.~Testa}\email{e.testa@ipnl.in2p3.fr}
\author{M.~Bajard}
\author{M.~Chevallier}
\author{D.~Dauvergne}
\author{F.~Le~Foulher}
\author{J.-C.~Poizat}
\author{C.~Ray}
\author{M.~Testa}
\affiliation{Universit\'e de Lyon, F-69622, Lyon, France;\\ Universit\'e Lyon 1, Villeurbanne; CNRS/IN2P3, UMR5822, IPNL}
\author{N.~Freud}
\author{J.M.~L\'etang}
\affiliation{CNDRI (nondestructive testing using ionizing radiation) laboratory, INSA-Lyon,  69621 Villeurbanne cedex, France}

\begin{abstract}
By means of a time-of-flight technique, we measured the longitudinal profile of prompt $\gamma$-rays emitted by 73 MeV/u $^{13}C$  ions irradiating a PMMA target. This technique allowed us to minimize the shielding against neutrons and scattered $\gamma$-rays, and to correlate prompt gamma emission to the ion path. This correlation, together with a high counting rate, paves the way toward real-time monitoring of the longitudinal dose profile during ion therapy treatments. Moreover, the time correlation between the prompt gamma detection and the transverse position of the incident ions measured by a beam monitor can provide real-time 3D control of the irradiation.

\end{abstract}

\maketitle

When interacting with matter, fast protons or light ions  deposit a maximum energy density in a region localized at the end of their path (the Bragg peak). The longitudinal and lateral dispersions of beams with energies above tens of MeV/u remain small, even for ranges of several centimeters. Moreover, at the Bragg peak, the biological effectiveness, {\it i.e.} the ability of ionizing projectiles to induce cellular death, is much higher for ions than for photons. This makes ion therapy highly competitive in the treatment of tumours, thus accounting for the recent worldwide development of medical ion accelerators. Another particularity of hadrontherapy originates in the important part played by nuclear reactions. On the one hand, such reactions induce a dispersion of the deposited dose by nuclear fragments and the associated radiations \cite{endo07,gun04}. On the other hand, the {\it in situ} control  of the dose deposited by the primary beam is made possible by the detection of radiations from nuclear reactions. For instance, Positron Emission Tomography (PET) is used to determine the annihilation distributions of $\beta^+$ particles emitted by the radionuclides activated during irradiation \cite{engh04}. After irradiation, these distributions are compared to those calculated by dedicated Monte Carlo simulations in order to control the treatment. Not only PET, but also prompt-$\gamma$ radiation detection may be regarded as a promising technique to monitor (in real time) the dose during ion irradiation treatments, since, in much less than a nanosecond following the ion beam impact, $\gamma$-rays and neutrons are emitted by excited nuclei with a high probability every time that nuclear fragmentation occurs. Thus prompt radiation may be used to monitor the dose inside a patient, provided the fragmentation probability along the path is clearly known. Moreover, the emission profile of prompt radiation is closely correlated to the primary beam range, since fragmentation occurs all along this path. Thus, it would be possible not only to measure the total dose, but also to determine its longitudinal distribution. The monitoring of hadrontherapy obviously requires 3D cartography of the dose, which could be obtained quite easily by a time correlation between the $\gamma$-ray detection and the determination of each incident ion transverse position.
Technically, $\gamma$-prompt detection  requires that radiations coming directly from the ion track be discriminated from those scattered in the surrounding matter, namely elastically- and inelastically-scattered neutrons and Compton-scattered $\gamma$-rays. Recently, Min {\it et al.} studied the longitudinal distribution profile of prompt $\gamma$-rays resulting from target nuclei fragmentation induced by proton impacts on thick targets, using a collimated scanner detection set-up. They found that  information on the ion range can be obtained for proton energies up to 200 MeV~\cite{min06}. 
In the present study we propose to extend this work to carbon ion beams, which involves both target- and projectile-nuclei fragmentations. We improve the detection technique by using combined time-of-flight and energy discrimination techniques. More importantly, we obtain an estimate of the measured $\gamma$-ray yield, proving the feasibility of the method for the on-line monitoring of irradiations.

Simulation codes, such as Geant4~\cite{geant4}, can provide useful figures in this case. One may expect an inelastic collision rate of 5-10\% for incident 73 MeV/u carbon ions irradiating a plastic target, with a typical emission rate of approximately two $\gamma$ photons and three neutrons for each reaction. The energy spectrum of photons ranges from a few tens of keV up to roughly 20 MeV, and neutron velocities from zero to approximately the projectile velocity, {\it i.e.} about one third of the speed of light. Thus, using a $\gamma$-ray detector at a sufficiently large distance from the target it is possible to distinguish clearly between $\gamma$ photons and neutrons by means of a time-of-flight technique.

\begin{figure}[hbtp]
\includegraphics[width=1\textwidth]{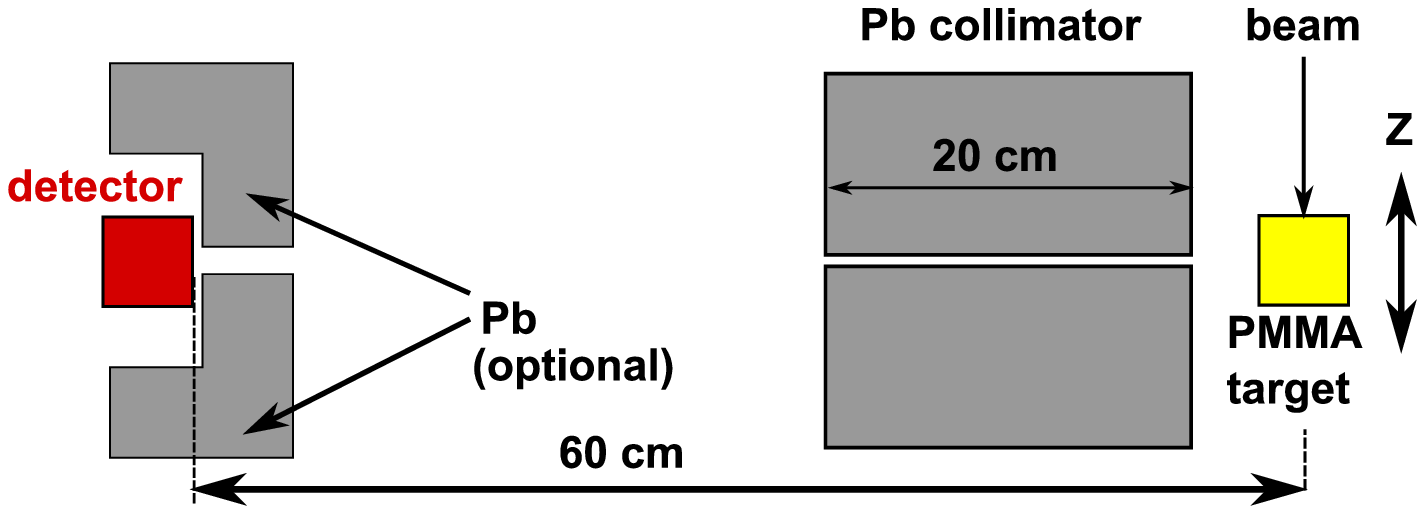}
\caption{\label{Setup} Diagram of the experimental setup.}
\end{figure}

The experiment, performed at the GANIL facility (Caen, France), used a beam of 73 MeV/u $^{13}C^{6+}$ ions. The $^{13}C$ and $^{12}C$ ions present the same fragmentation cross sections and the same yield of gamma-ray emission within 10\%. The use of  $^{13}C$ only leads to an increase in the neutron emission yield corresponding to the additional neutron ~\cite{geant4}. The beam was pulsed (beam pulses of $\sim$1 ns every 80 ns), and the intensity on the target was set to approximately 1 nA in order to minimize the acquisition time. Figure \ref{Setup} shows a diagram of the detection set-up. The beam impinged on a PMMA ($C_5H_8O_2$) cubic target (10 cm edge) which could be translated along the direction of the beam.

The main detector, a cylindrical NaI(Tl) detector of 5 cm in diameter and length, was placed at $90\degree$ to the beam direction behind a 20 cm thick lead collimator, with the possibility of adding shielding around the detector according to the tightness of the collimation. The remote target translation allowed us to select the part of the target to be viewed by the collimator.
The total dose was monitored with a second NaI(Tl) detector with no collimator, placed at a greater distance ($\sim$ 1 m) in order to obtain a counting rate proportional to the beam intensity, almost irrelevant of the target position. This monitor was calibrated with a Faraday cup at higher intensities. A key point of our experiment was the discrimination between detected $\gamma$-rays and neutrons by the time-of-flight technique: the difference $T$ between the detection time by the $\gamma$-ray detector and a given phase of the RF-signal from the accelerator was measured using a time-to-analog converter. Finally, the deposited energy $E$ and the time $T$ associated to each detected radiation were recorded, along with the number of counts registered by the dose monitor during a run.

On-line monitoring of the dose and position of the Bragg peak during hadrontherapy treatment requires  high counting rates and a comfortable signal-to-background ratio, together with a spatial resolution in the order of a few millimeters. Thus, in order to obtain a distribution of prompt radiations that meets these requirements, both the detection and shielding geometries and the off-line event selections were tuned as follows: i) the target-to-main detector distance and the collimator slit were set to 60 cm and 2 mm respectively; ii) off-line, the detected radiations were selected as a function of the energy they deposited in the detector and of their time-of-flight.
In fig.2 we show time-of-flight spectra and, in inset, a typical energy spectrum. The shape of energy spectra is relatively independent of the geometry of the set-up. The only visible peak, at 511 keV, is mainly due to pair creation by high energy photons, followed by positron annihilation. Other peaks corresponding to detector activation or natural background could be observed with high resolution detectors such as germanium detectors \cite{paro05}. Nevertheless, the observed energy spectrum is dominated by a continuous component corresponding to gamma-rays emitted before equilibrium is reached in highly excited matter \cite{nife90}. It is to be noted that an acquisition time of a few minutes (the order of the lifetime of the main $\beta^+$ emitters) is necessary to observe the annihilation of $\beta^+$ emitted by radionuclides (mainly $^{11}C$), and that this acquisition has to be started about one hundred nanoseconds after the beam impact to eliminate prompt radiation. Because of these constraints, the PET acquisition used to control hadrontherapy is performed during the irradiation between the beam extraction bunches \cite{cres05} and prolonged for a few minutes after the treatment.

\begin{figure}[hbtp]
\includegraphics[width=1\textwidth]{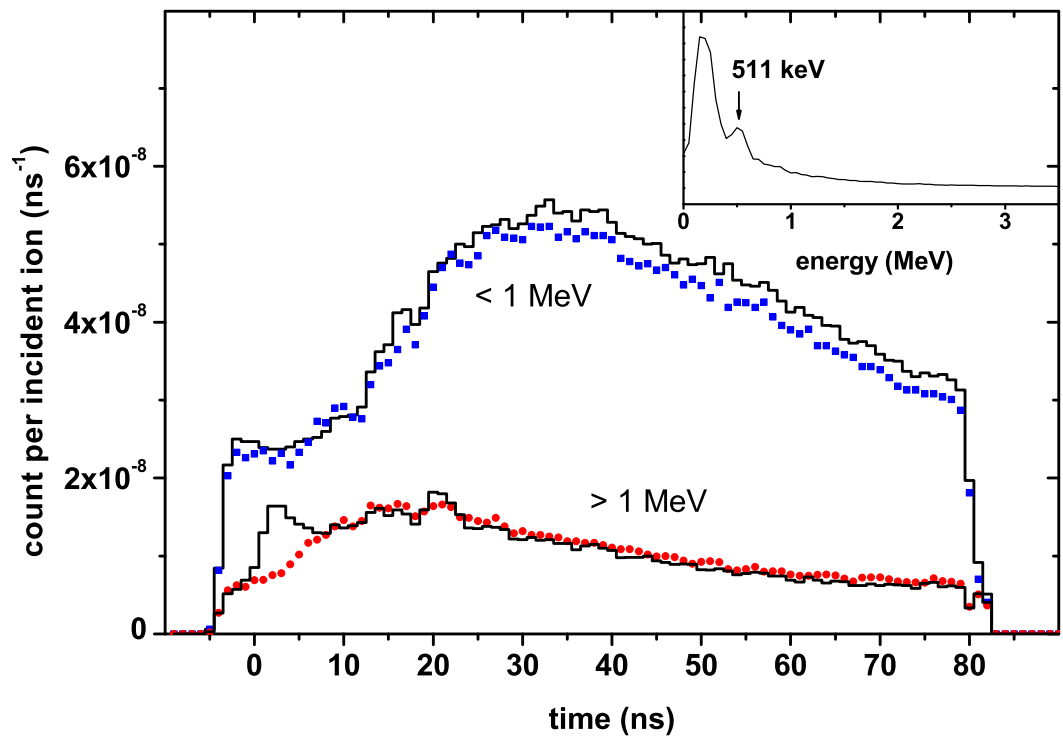}
\caption{\label{TAC} Time-of-flight spectra measured for two target locations $Z$ and for two selections on energy $E$ deposited in scintillator : $Z=8$~mm (lines), $Z=26$~mm (dots), $E<1$~MeV and $E>1$~MeV. Origin of time scale: ion impact on target. Inset: measured energy spectrum.}
\end{figure}

\begin{figure}[hbtp]
\includegraphics[width=\textwidth]{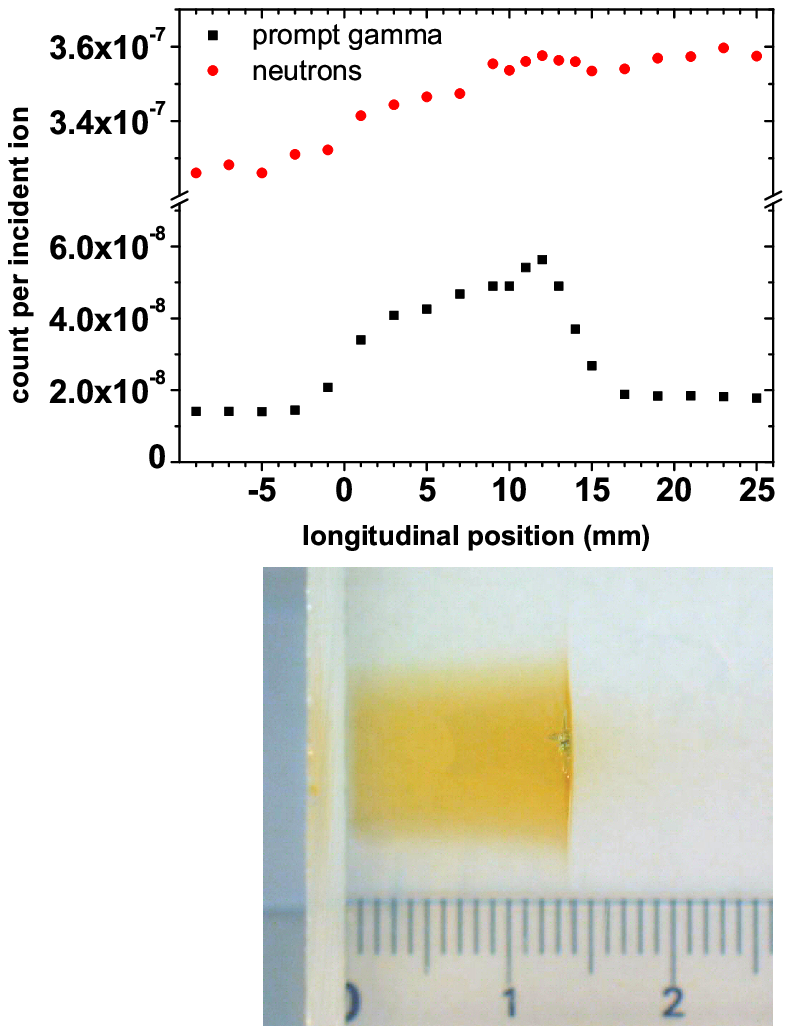}
\caption{\label{Rendement} Upper part: detection rates ($E>1$~MeV) as a function of longitudinal position of target, obtained for two different  time-of-flight ranges:   $2 < T < 10$~ns (prompt $\gamma$-rays, squares) and  $T > 10$~ns (neutrons, circles). Bottom image: scaled photograph of irradiated PMMA sample.}
\end{figure}

While the shape of the energy spectrum is virtually independent of the geometry of the set-up, the time-of-flight spectrum depends strongly on the target-to-detector distance and on the part of the ion path observed in the target. Figure \ref{TAC} presents the time-of-flight spectra measured when the collimated detector views two zones of the target (respectively in the middle of the ion path and 12 mm beyond the Bragg peak region) for two selections of energy $E$ deposited in the scintillator (above and below 1 MeV). If we first consider the time-of-flight spectrum obtained with the energy selection $E>1$ MeV, for both Z target positions we observe, as expected, a broad time-of-flight distribution due to detected massive particles (mainly neutrons). When the detector is aimed at a region within the ion path, we observe an additional narrow component corresponding to prompt radiation, {\it i.e.} to photons. For this target position and with this selection of the measured energy deposition, the value of the peak-to-continuum ratio is approximately one. With the energy selection $E<1$ MeV, the time-of-flight distributions measured at the two locations are very similar, and are completely dominated by neutrons.Figure~\ref{Rendement} gives the detection rates as a function of the longitudinal position $Z$ obtained for two different time-of-flight selections: the prompt gamma peak and a neutron selection. We selected energies $E>1$ MeV in both cases: this selection corresponds to the best compromise between the counting statistics of the events selected and the signal-to-background ratio. The ion range was determined by measuring the length of the darkened area of the irradiated target (see photograph of an irradiated sample at the bottom of figure \ref{Rendement}. The dose accumulated by this sample is quite high ($>10^5$ Gy), and thus the damage is clearly visible. The yellowish damaged zone is darker at the Bragg peak location, with a very sharp edge at 14 mm. Up to 1 cm beyond this edge, it is possible to distinguish an extension of damage due to light particles. When the profiles in the two images are compared, it can be observed that the prompt gamma yield is strongly correlated to the ion path in the target, whereas the neutron detection profile is almost flat (the lead collimator does not allow the filtration of neutrons), and that the highest signal-to-background ratio is around 3 at the Bragg peak location. One of the striking features of the photon detection profile is, in fact, the enhancement observed when the detector is targeted on the Bragg peak location. This can be partly attributed to a possible increase of fragmentation cross sections when the ion energy decreases~\cite{she89} and, more probably, to the effect of relatively long radiative decay times of excited fragments. Simulations have shown that the observed profile could be reproduced assuming an average radiative decay time of a few tens of picoseconds, allowing a large fraction of the excited fragments to halt before emitting a $\gamma$ photon. 

This study shows that measured prompt radiation yields are extremely promising with a view to designing an on-line monitor of the ion path during ion therapy, since, with our detection set-up, we observed approximately $10^{-7}$ prompt gamma per carbon ion for a range of 14 mm in PMMA. In the present case, the fraction of the azimuth covered by the detector is only 1.3\%, and only one angle of observation (perpendicular to the beam direction) is used. The solid angle could therefore be increased by two orders of magnitude, and higher counting rates could be obtained by observing a larger zone of the target, since the geometry of the  collimator used here aimed at optimized spatial resolution at the Bragg peak location with a tight collimation. Observation at several angles with respect to the beam direction may also increase the solid angle of detection considerably. The intrinsic efficiency of the detector can also be improved, as well as the discrimination between neutrons and photons (NaI(Tl) detectors are not the best suited). Depending on tumour depth, between $10^6$ and $10^7$ ions are required to treat a $1$ cm$^3$ tumour during one irradiation session delivering a 1-2 Gy equivalent dose~\cite{sch07}, and thus on-line control with prompt radiation seems realistic.
The present study also demonstrates that time-of-flight selection is effective in discriminating $\gamma$-rays from neutrons, which, in turn, makes radiation shielding requirements less stringent, and allows high detection efficiency. The time structure of the GANIL pulsed beam is an ideal case. For other kinds of beams, for example ion beams delivered by synchrotrons with relatively long pulses of typically 80 ns, the TOF technique makes it possible to decrease the integrated background proportionally to the ratio between the prompt gamma time window and the average time separating two consecutive incident ions. In this case, the time reference should be given by a segmented detector (e.g. a hodoscope), which could, moreover, enable the monitoring of 3D dose cartography by coupling its measurement of the incident ion transverse position with our experimental set-up. 

Within the framework of the PNRH (French Hadrontherapy National Research Programme), the GDR MI2B research network (supported by the French CNRS), and the ETOILE Center for Hadrontherapy, we are now undertaking further studies in order to improve this promising technique of dose monitoring.

\newpage

\end{document}